# Graded nanocomposite metamaterials for a double-sided radiative cooling architecture with a record breaking cooling power density

Lyu Zhou[*, 1], Haomin Song[*, 1], Nan Zhang[1], Jacob Rada[1], Matthew Singer[1], Huafan Zhang[2], Boon S. Ooi[2], Zongfu Yu[3], Qiaoqiang Gan[†, 1]

1 *Department of Electrical Engineering, The State University of New York at Buffalo, Buffalo, NY 14260, USA.*

2 *KAUST Nanophotonics Lab, King Abdullah University of Science and Technology, Thuwal 23955-6900, Saudi Arabia*

3 *Department of Electrical and Computer Engineering, University of Wisconsin, Madison, Wisconsin 53705, USA*

\* These authors contributed equally to this work.

† Email: qqgan@buffalo.edu

**Abstract-** As an emerging electricity-free cooling technology, radiative cooling employs outer space as the heat sink. With this, a sky-facing thermal emitter is usually required. Due to the blackbody radiation limit at ambient temperature, the maximum cooling power density for a single-faced radiative cooling device is ~156.9 W/m$^2$. Here we report a double-sided radiative cooling architecture using graded nanocomposite metamaterials (GNM) designed for a vertically aligned thermal emitter. This GNM structure possesses an optical absorption of over 90% throughout the solar spectrum, and exceeds 90% reflection in the mid-infrared spectral region. With this configuration, both sides of a planar thermal emitter can be used to perform radiative cooling and a record cooling power density beyond 280 W/m$^2$ was realized in a single thin-film thermal emitter. Under the standard pressure, we realized a temperature reduction of 14 °C below the ambient temperature in the laboratory environment, and over 12 °C in the outdoor test.

Electricity-driven cooling is one of the major end-uses of energy that is responsible for global peak electricity demand. For instance, in tropic areas, air conditioning consumes ~70% of the total energy consumption used by buildings [1]. Of the fuel used in vehicles, ~29% is used for cooling and 25%-33% is lost due to heat dissipation [2]. During the summer, the temperature inside the car parked in direct sunlight can reach 50 ºC, which is extremely uncomfortable and can be dangerous to humans and pets [3]. Therefore, a passive cooling strategy that cools without electricity could significantly impact global energy consumption as well as safety. Sky cooling is an emerging electricity-free cooling technology using outer space as the heat sink. However, the maximum cooling power for a single-faced radiative cooling device is ~160 W/m$^2$ due to the blackbody radiation limit at ambient temperature. Here we report a double-sided radiative cooling architecture using graded nanocomposite metamaterials (GNM) designed for a vertically aligned thermal emitter and demonstrate a record cooling power density beyond 280 W/m$^2$ in a single thin-film thermal emitter. Importantly, this strategy did not simply waste the solar input energy. It can keep the solar thermal and radiative cooling effects in a single system, enabling the development of hybrid 2-in-1 solar heating and sky cooling utilities.

Due to the transparency of the Earth's atmosphere in the wavelength range of 8-13 μm, objects can emit their heat through this spectral window and realize a cooling effect. Building upon this heat exchange channel, sky-cooling emerges as an electricity-free cooling technology. However, there are two major challenges in order to implement this electricity-free cooling technology in a practical environment: (1) the object has to have direct access to the sky (i.e., with direct sky-facing emitting surfaces), and (2) one needs a solar transparent material/structure with thermal emission energy stronger than its optical absorption of the solar input energy for day-time cooling [4]. In recent years, many groups have been attempting to discover radiative cooling effects in biomaterials (e.g. [5, 6]) and develop high performance and low cost thermally emissive metamaterials (e.g. [7-13]). Although various advanced thermal photonic materials with different spectral selection features have been reported, the average measured cooling power density of these devices is ~100 W/m$^2$ during the day, and ~120 W/m$^2$ during the night [14-17]. These reported values can be attributed to the intrinsic optical and thermal characteristics of the system [18-19] or material [20-26] as well as the actual weather conditions [21-22, 27-28].

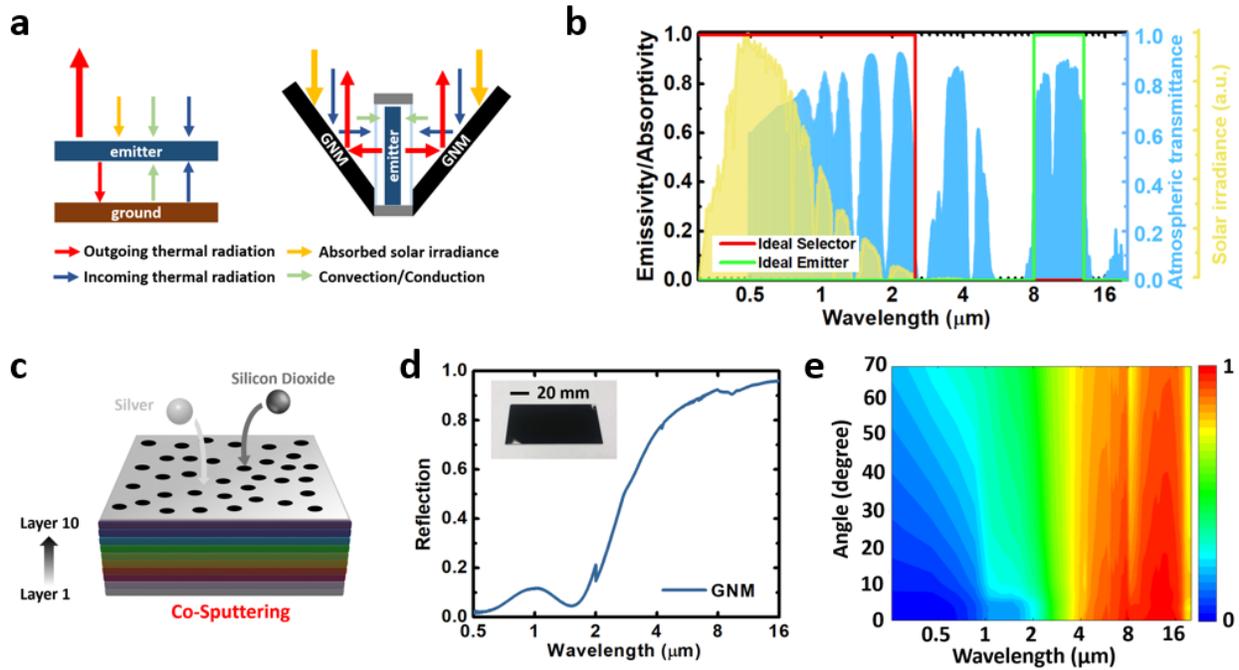

**Figure 1| GNM spectral selector.** (a) Diagrams of the power tradeoff in a conventional single-sided radiative cooling system (left panel) and a double-sided system (right panel), respectively. The red arrow indicates the outgoing thermal radiation from emitter, the blue arrow indicates the incoming thermal radiation from both atmosphere and surrounding objects, the green arrow indicates the non-radiative power exchange, and the yellow arrow indicates the absorbed solar irradiation. (b) Emissivity of the ideal cases: the red line indicates an ideal selector and the green line indicates an ideal emitter for daytime radiative cooling design. The AM 1.5 solar spectrum (yellow) and the atmospheric transmittance (blue) are plotted for reference. (c) Schematic of the GNM deposition sequence. Ten layers of Ag-SiO$_2$ nanocomposite were deposited with different mixture ratios. (d) The measured reflection spectrum of the GNM plate. The inset shows a photograph of a GNM plate. (e) Measured angular-dependent reflection spectra of the GNM plate.

For a planar thin film thermal emitter, both the top and bottom surfaces emit thermal radiation. However, only the sky-facing surface causes a noticeable cooling effect, as reported in previous radiative cooling experiments (e.g. [4, 6-13, 18-26]). The bottom surface can only exchange the thermal radiation with the ground (left panel in **Fig. 1a**). Therefore, a waveguide strategy that can guide the thermal emission from the two surfaces to the sky will break the cooling power limit of the single-sided thermal emitter. In this work, we report a graded nanocomposite metamaterial (GNM) spectral selector to function as a thermal radiation waveguide/mirror, as well as a solar absorber. This structure is capable of absorbing over 90% of electromagnetic radiation in the solar spectrum, while also maintaining a reflection of over 90% in the mid-infrared region, including 8-13 μm. When a vertical thermal emitter is coupled with this spectral selector (right panel in **Fig. 1a**), its thermal emission can be mostly reflected to the sky while the solar input is absorbed by the thermal selector simultaneously. As a merit, both sides of a planar thermal emitter can be used to emit heat and we realize a record high cooling power density of over 280 W/m$^2$ in a laboratory environment, twice as much as the previously reported cooling power density from the single sky-facing surfaces [7, 10, 18]. Without special vacuum thermal isolation, we realized a temperature reduction of 14 ºC below the ambient temperature in the laboratory environment and over 12 ºC in the outdoor test under standard atmospheric pressure. In addition, the spectral selectivity of GNM plates enables an efficient solar absorption and restrains thermal radiation, resulting in a tremendous temperature rising in GNM plates. Such simultaneously solar heating and radiative cooling configuration paves the way towards wider applications in hybrid solar heating and radiative cooling utilities.

In principle, to realize daytime radiative cooling, the emitter is required to be strongly absorptive in mid-infrared wavelength range while has minimized absorption in solar wavelength range (up to 2.5 μm, as shown in Fig. 1b). Here we employ an ideal spectral selector to realize this spectral selectivity: it efficiently absorbs solar light and simultaneously reflects thermal radiation from emitter (as shown by the red curve in **Fig. 1b**). In this experiment, 10 layers of co-sputtered Ag-SiO$_2$ nanocomposite films were stacked to construct the GNM film on a glass substrate, labeled as Layer 1 to Layer 10 in **Fig. 1c** (see fabrication details in *Materials and Methods* section) [29-32]. The nominal thickness of each layer is ~30 nm. As a result, a strong solar absorber was realized. As shown by the inset of **Fig. 1d**, the GNM film is visibly black. Its average reflectivity from visible to infrared (up to 2000 nm) is below 8%. Intriguingly, its reflection increases rapidly and remains over 90% throughout the mid-IR spectral range, covering the atmospheric window of 8-13 μm (see its infrared image in *Fig. S1*). Remarkably, this superior spectral selection feature can be retained over a broad incident angle range, as shown in **Fig. 1e**: within the incident angle of 50°, the absorption is over 90% in solar spectral regime and the reflection is over 95% in the 8-13 μm range. To reveal the mechanism of this spectrally selective GNM film, we first analyze its microscopic features.

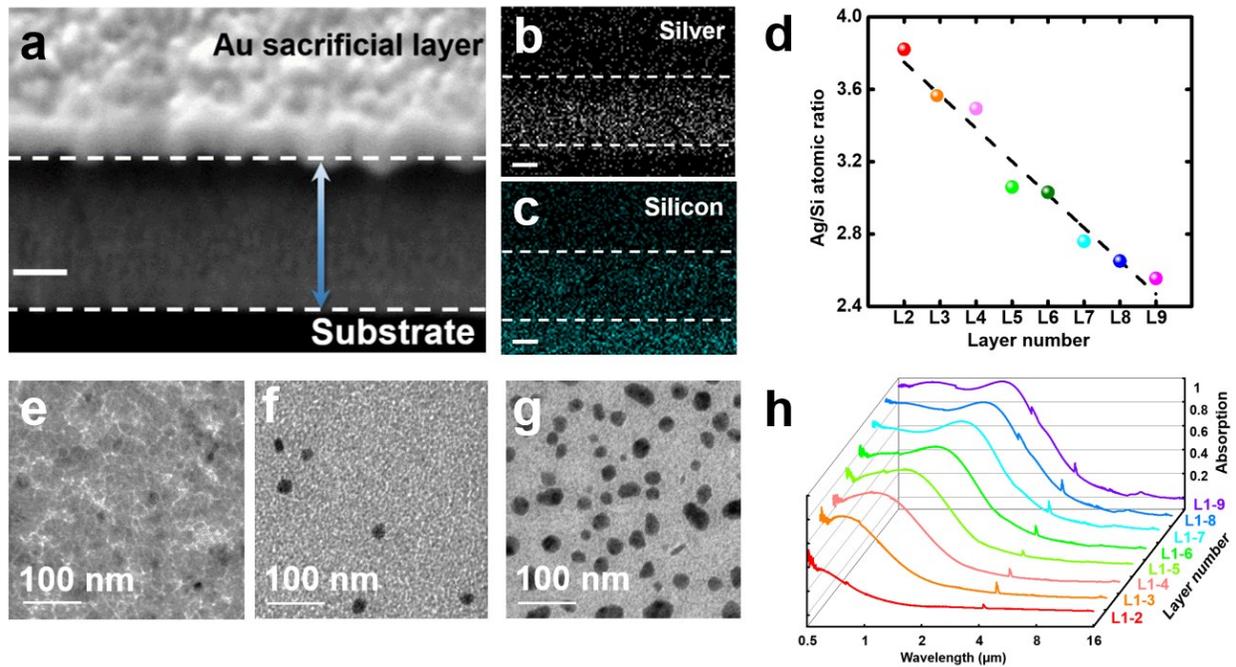

**Figure 2| Microscopic characterization of the GNM film.** (a) SEM image of the cross section of a GNM plate. The arrow indicates the GNM region. The top Au layer is the sacrificial layer for SEM imaging. (b)-(c) EDS element mapping images for (b) Ag and (c) Si, respectively. (d) Element atomic ratio of Ag/Si for the reference sample of each layer (Layer 2 to Layer 9). (e)-(g) TEM images of three reference samples: (e) Layer 8, (f) Layer 5 and (g) Layer 3. (h) Absorption spectra of GNM samples with different numbers of stacked layers.

**Fig. 2a** shows the cross-sectional scanning electron microscopic (SEM) image of the GNM film (indicated by the arrow), showing the graded profile of the Ag-SiO$_2$ composite. We performed the energy dispersive X-Ray spectroscopy (EDS) analysis on the GNM film, showing the graded distribution of Ag (**Fig. 2b**) and Si elements (**Fig. 2c**), respectively. To better reveal their graded distributions, we fabricated reference samples for each layer of the graded Ag-SiO$_2$ nanocomposite film and analyzed their Ag/Si atomic ratio in **Fig. 2d** (see details in *Materials and Methods* section). One can clearly see the monotonically decreased Ag/Si atomic ratio. The transmission electron microscopy (TEM) characterization was also performed to further reveal the microscopic morphology of the GNM film. As shown in the TEM images for three layers (Layer 8, Layer 5 and Layer 3, respectively), **Figs. 2e-2g**, Ag particles with dimensions of 20-40 nm were embedded in the continuous SiO$_2$ film. To further reveal the optical feature of these graded index films, we performed a series of absorption characterization on GNM films with different stacked layers, as shown in **Fig. 2h**. One can see that as the number of layers increased, the optical absorption in the visible to near-infrared spectral range will improve accordingly, as a result of the reduced impedance mismatch of the dielectric constant. This GNM film is able to simultaneously, and efficiently, absorb solar illumination and reflect thermal radiation. This will enable its use as a spectral selector for direction-guided radiative cooling.

Using these GNM plates, we built a V-shaped mirror structure and measured its radiative cooling performance. As shown in **Fig. 3a**, a thin film thermal emitter was placed in a polystyrene box with two open sides exposed to two GNM mirrors. The box was then sealed with polyethylene film. As a result, thermal radiation from both sides of the emitter was directed to the remote heat sink (i.e., the sky). By tuning the angle of these two GNM plates, the output thermal emission can be optimized (see *Fig. S2*). As shown in **Fig. 3b**, we modeled the angular distribution of the thermal radiation at the wavelength of 10 μm when the GNM plates were tuned to 45°. Compared with the wide thermal radiation pattern from a conventional single-sided system (the blue curve), most of the thermal emission reflected by the V-shaped structure is confined within ±39.75°(the red curve), enabling a beam-coupling effect to the thermal radiation [22,27]. Moreover, on merit of the spectral selective property of the GNM mirror, this V-shaped configuration can direct thermal radiation from both sides of a standing planar emitter, achieving a record-breaking radiative cooling power density. To validate this hypothesis, we then performed several field-test in both laboratory and outdoor environment.

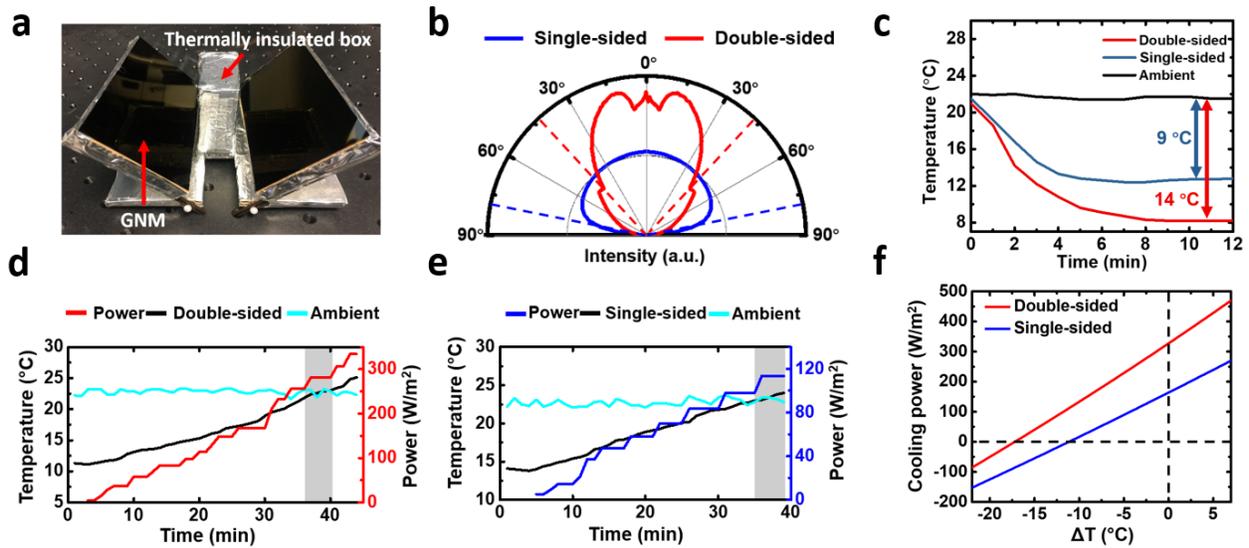

**Figure 3| Radiative cooling performance of the double-sided thermal emitter system.** (a) A photo of the double-sided system made of a thermally insulated box with a V-shaped GNM plates. (b) The modeled angular emissive intensity of the single-sided (blue curve) system and V-shaped double-sided system (red curve). The dashed curve shows the angle at which the emitted intensity reaches 50% of its corresponding maximum. (c) Temperature measurement of the single-sided system (blue curve) and the double-sided system (red curve) in the laboratory environment. The black line shows the ambient temperature. (d)-(e) Measured heater power and temperatures of the emitter in the (d) double-sided system and (e) single-sided system. The black and cyan curves indicate the temperature of the emitter and the ambient temperature, respectively. The shadow areas indicate the moment when the emitter's temperature matches the ambient temperature. The heating powers (red and blue curves in (d) and (e), respectively) in these two regions reflect the cooling powers of the double-sided and single-sided systems, respectively. (f) Estimated cooling powers of the two systems as a function of the temperature difference ($\Delta T$) between the emitter and ambient temperature.

First, liquid nitrogen was employed as the cold source to simulate indoor radiative cooling test (similar to our previous work [22]). For this experiment, black aluminum was used as the tested emitter due to its near unity emissivity over the IR spectrum range (see its optical emissivity spectrum in *Fig. S3*). As shown in **Fig. 3c**, the temperature of the double-sided system was reduced by ~14 °C, ~5°C lower than the control system (i.e., ~9.0°C). To experimentally reveal cooling power, we employed a heat patch attached to the black aluminum to heat it to the ambient temperature (following the procedure reported by ref. [18]). As shown in **Figs. 3d** and **3e**, the surface temperatures of the black aluminum in the double-sided system and single-sided system were plotted by black curves. When they reached the ambient temperature (i.e., the cyan curves), the measured cooling powers are ~280.7 W/m² for the double-sided system (see the red curve in the shaded region in Fig. 3d) and 113.7 W/m² for the single-sided system (see the blue curve in the shaded region in Fig. 3e). To further evaluate the measured cooling performance compared with the theoretical upper limit, the net cooling power density ($P_{net}$) of an ideal blackbody emitter was calculated as a function of the temperature difference, $\Delta T$, using equation (1):

$$P_{net} = P_{rad}(T_{dev}) - P_{amb}(T_{amb}) - P_{obj}(T_{obj}) - P_{nonrad}(T_{dev}, T_{amb}), \qquad (1)$$

Here $P_{rad}(T_{dev})$ is the outgoing radiation power from the thermal emitter at the temperature of $T_{dev}$, $P_{amb}(T_{amb})$ is the incoming radiation power from the ambient environment at the temperature of $T_{amb}$, $P_{obj}(T_{obj})$ is the incoming radiation power from surrounding objects (i.e., liquid nitrogen and GNM plates) at the temperature of $T_{obj}$, and $P_{nonrad}(T_{dev}, T_{amb})$ is the non-radiative power transferred to the emitter through convection and conduction (see calculation details in **Materials and Methods**). As shown in **Fig. 3f**, at an ambient temperature of 22 °C and $\Delta T = 0$ (indicated by the vertical dashed line), the cooling power of the double-sided system reaches ~313.8 W/m², twice as much as that of the single-sided system (i.e., ~156.9 W/m²). And when the emitters reach thermal equilibrium (see the horizontal dashed line in Fig. 3f), they can realize a temperature reduction of ~16.9 °C in the double-sided system, and ~10.8 °C in the single-sided system, respectively. Therefore, the indoor experimental results are proved to be very close to the theoretical prediction without taking solar input into consideration.

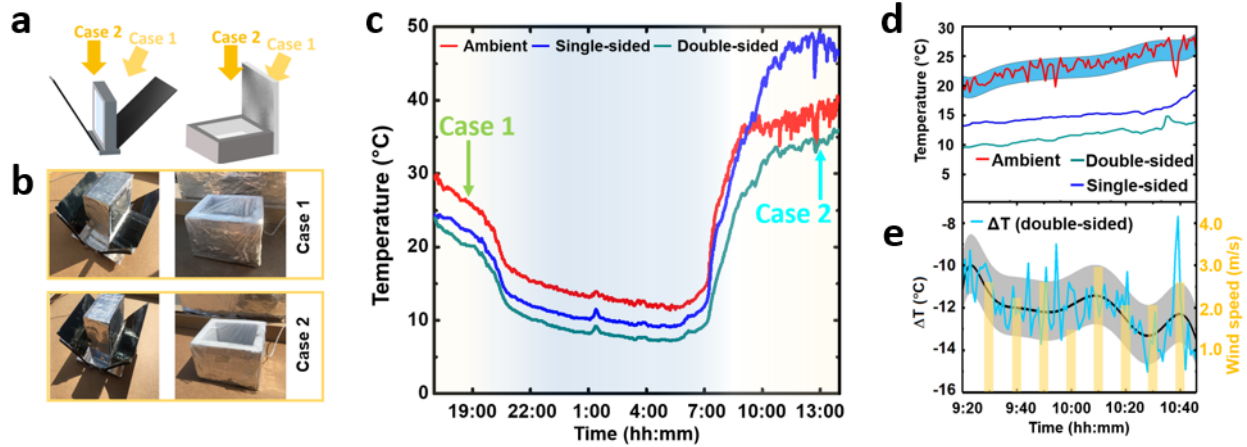

**Figure 4| Improved day-time sky cooling.** (a) Schematic of the daytime radiative cooling tests in Buffalo, NY. (b) Photos of the outdoor radiative cooling measurement of case 1 (upper panel) and case 2 (lower panel) with different solar incident angles. (c) Measured temperature of the emitter in the double-sided system (green curve) and the single-sided system (blue curve). The ambient temperature is shown by the red curve. (d) The best cooling performance was measured on May 21, 2019 at Buffalo, NY. (e) The temperature difference between the ambient temperature and the emitter in the double-sided system. The yellow bar shows the wind speed at the test location.

To demonstrate the superior cooling capability of the proposed double-sided emitter structure with GNM plates, we performed outdoor measurements as well in Buffalo, NY. A single-sided control experiment was performed simultaneously. In these two systems, solar-transparent PDMS films were anchored at the center of polystyrene boxes to replace black aluminum in previous session as the thermal emitter (**Fig. 4a**). Photographs of the experimental setup are shown in **Fig. 4b**: a highly reflective aluminum board was used to shelter the control system from sun light (similar to the one used in ref. [19, 22]), while the double-sided emitter system was directly exposed to the sun. We performed a 24-hour continuous outdoor test in Buffalo, NY on Aug. 4-Aug. 5, 2019 (with a humidity of 30% in the day and 95% at night, see more discussion of the weather conditions in *Fig. S4*). Due to the vertically anchored thermal emitter in the double-sided system, the cooling performance of the system is very different from the single-sided system. As shown by Case 1 in Fig. 4b, in the single-sided system, the incident sun light was blocked by the shelter, but directly illuminate the PDMS film in the double-sided system. As a result, the cooling performance of the double-sided system degraded significantly. Nevertheless, due to fact that PDMS is highly transparent in solar wavelength range, the estimated absorbed solar energy by PDMS film (~113.95 W/m$^2$) was still much lower than the net radiation power (>200 W/m$^2$). As a result, a sub-ambient cooling was still obtained in the double-sided configuration without the implementation of expensive sun tracking systems (e.g. [15]). One can see that at 18:44 in Fig. 4c (the green arrow), the double-sided system realized a temperature reduction of 6.2 ºC, while the ΔT in the single-sided system was 4.2 ºC.

As the solar input intensity steadily rose over the course of the day/afternoon (i.e., Case 2 in Fig. 4b), direct illumination was blocked by the side wall of the polystyrene box in the double-sided system, while the planar shelter cannot block the normal incidence in single-sided system, especially in places that receive large amount of harsh sun (e.g. Saudi Arabia, as shown in *Fig. S5*). Therefore, the cooling performance of the double-sided emitter system is much more superior to that of the single-sided system, especially under the normal solar incidence when air-conditioning is the most demanding! As shown in Fig. 4c, the double-sided system remained ~4.5 ºC lower than the ambient temperature at 12:58 (the cyan arrow). In contrast, the temperature inside the single-sided system is ~10 ºC higher. By reviewing the overall performance of the two systems shown in Fig. 4c, the double-sided system outperformed the single-sided one in an all-day sub-ambient cooling.

As previous noted, the radiative cooling performance is heavily dependent on weather conditions. During May - Aug. 2019, we performed over 70-hours outdoor experiments on selected sunny days (see details in *Fig. S6*). On May 21, 2019 (with a very clear sky and humidity of ~17% during the experiment time), an even more intriguing performance was observed, as shown by the red curve in **Fig. 4d**. Obvious fluctuations were observed in the ambient temperature due to the relatively strong wind. To better reveal the reduced temperature, we plotted the ΔT curve of the double-sided system in **Fig. 4e**. From 9:20 to 11:00, the average temperature reduction of the double-sided emitter system was 12±1.2 ˚C, corresponding to a cooling power density of 261.5 W/m$^2$. The shaded single-sided system, employed as comparison, however, only realized a temperature reduction of 8±1.5 °C, corresponding to a cooling power density of 102.3 W/m$^2$ (similar to our previously reported results [22]). This was a record high temperature reduction for this simple system with no special thermal insulation (e.g. [19]).

In conclusion, we developed a GNM that can selectively absorb solar light and reflect mid-infrared light efficiently. Using such GNM plates, a V-shaped architecture was built to couple the thermal radiation from both sides of the planer emitter to the sky, and realize a record high cooling power density of over 280 W/m$^2$. Under standard atmospheric pressure, we demonstrated a record temperature reduction of 14 °C in a laboratory environment, and 12±1.2 °C in the outdoor test. Moreover, in recently reported radiative cooling works, solar input and its heating effect were always undesired to the cooling system (i.e., scattered or reflected) [33-36]. To utilize this formerly wasted energy, the proposed GNM plate allows solar heating and radiative cooling to occur within the same system: according to our measurement, between 9:00 and 11:00, the temperature of the unsealed GNM plates was ~54 °C, given an irradiance of ~1010 W/m$^2$. Therefore, the double-sided emitter system can provide a 2-in-1 solution to integrate the cooling and heating into a single system perfectly: i.e., simultaneously providing a cold source (i.e., 10~15 °C) and a mild heat source (i.e., ~54 °C) for residential applications (e.g. the annual energy consumption for residential water heating accounts for 28% of all energy used in Los Angeles, California [37]). The hybrid solar-heating and radiative cooling system can contribute to a reduction in both cooling and heating costs, simultaneously [38-40], and can be integrated with thermoelectric generator systems to produce power [41].

**Materials and Methods**

**Fabrication and characterization of the GNM film.** The GNM film was fabricated in a Lesker Sputtering Coater System by co-sputtering Ag and $SiO_2$ simultaneously. A glass substrate was cleaned via standard wafer cleaning procedure. We started the deposition with a 30-nm-thick silver layer (i.e., Layer 1) followed by 8 layers of Ag/SiO$_2$ nanocomposites with gradually changing silver-to-silica ratio (Layer 2 to 9). Finally, a thin silica layer (~30 nm) was deposited as the top layer (i.e., Layer 10). This stacked structure formed a graded nanocomposite metamaterial with gradually changing refractive index, resulting in a broadband absorption from 300 nm to 2 μm [27]. On the other hand, its reflection in mid-infrared domain is high (Fig. 1d in the main text).

**Energy dispersive X-Ray spectroscopy (EDS) analysis of the GNM film.** During GNM film deposition, we fabricated reference samples for each individual layer of the 10 layers used in the GNM film. These films were also used in TEM characterization shown in Figs. 2e-2g in the main text. To reveal the graded distribution, we analyzed the element ratio of each layer. Since Layer 1 is pure silver and Layer 10 is pure silica, we only analyzed layers 2 to 9 using EDS to characterize the atomic ratio of all elements (Supplementary Table 1). The extracted data are plotted in Fig. 2d with a linear fit line to demonstrate the gradual change in the Ag/Si ratio.

**Cooling power estimation.** Based on Plank's law, the cooling power is estimated using equation (1) in the main text: i.e.,

$$P_{net} = P_{rad}(T_{dev}) - P_{amb}(T_{amb}) - P_{obj}(T_{obj}) - P_{nonrad}(T_{dev}, T_{amb}) \qquad (1).$$

Here $P_{rad}(T_{dev})$ is the outgoing power from the emitter at the temperature of $T_{dev}$, $P_{amb}(T_{amb})$ is the absorbed thermal radiative power from the ambient air at the temperature of $T_{amb}$, $P_{obj}(T_{obj})$ is the incoming thermal radiative power from surrounding objects, i.e., liquid nitrogen or GNM films, at the temperature of $T_{obj}$, and $P_{nonrad}(T_{dev}, T_{amb})$ is the non-radiative power exchange due to conduction and convection.

In the single-sided system, the emitter is facing the cold source. Hence the power emitted from the emitter can be estimated by:

$$P_{rad}(T_{dev}) = A \int d\Omega \cos(\theta) \int d\lambda I_{BB}(T_{dev}) \varepsilon_{dev}(\theta, \lambda) \quad (S1),$$

where $\int d\Omega$ is the angular integral of the emitter over the space, $I_{BB}(T) = \frac{2hc^2}{\lambda^5} \frac{1}{\exp\left(\frac{hc}{\lambda k_B T}\right) - 1}$ is the spectral radiance of a blackbody at a temperature T, $\varepsilon_{dev}$ is the spectral emissivity of the emitter (here we replaced it with the measured absorptivity based on Kirchhoff's radiation law). In this estimation, a black aluminum foil is employed as the thermal emitter (see its absorption spectrum in Fig. S3). $h$ is Planck's constant, $k_B$ is the Boltzmann constant, $c$ is the speed of light, $\lambda$ is the wavelength, and $A$ is the area of the emitter. In this estimation, $A$ is normalized to be 1 m².

The power absorbed from the ambient air, $P_{amb}(T_{amb})$, and from the surrounding objects, i.e., black aluminum immersed in liquid nitrogen, $P_{lN2}$, are given by the following two equations, respectively:

$$P_{amb}(T_{amb}) = A \int d\Omega \cos(\alpha) \int d\lambda I_{BB}(T_{amb}) \varepsilon_{dev}(\theta, \lambda) \varepsilon_{air}(\alpha, \lambda) \quad (S2),$$

where the angular-dependent emissivity of atmosphere $\varepsilon_{air}(\alpha, \lambda)$ is given by $\varepsilon_{air}(\gamma, \lambda) = 1 - [1 - \varepsilon_{air}(0, \lambda)]^{\frac{1}{\cos\gamma}}$. Here, $\varepsilon_0(\beta, \lambda)$ is the emissivity of the black aluminum, $T_{amb}$ is the temperature of the ambient air.

The non-radiative power loss is introduced by conduction and convection, which is given by

$$P_{nonrad}(T_{dev}, T_{amb}) = q(T_{dev} - T_{amb}) \quad (S3).$$

Following the calculation procedure disclosed in previous works[19, 20], the cooling power is estimated under the precondition that the emitter temperature is equal to the ambient temperature, and the non-radiative power loss can hence be neglected. In this estimation, the single-sided cooling power at 295 K is 156.9 W/m².

The radiative heat transfer between the GNM film and the emitter surface is also estimated as a function of GNM temperature using eq. S4.

$$P_{obj}(T_{obj}) = \frac{\sigma(T_{obj}^4 - T_{dev}^4)}{\frac{1-\epsilon_{obj}}{A_{obj}\epsilon_{obj}} + \frac{1-\epsilon_{dev}}{A\epsilon_{dev}} + \frac{1}{A_{obj}F}} \qquad (S4).$$

The value F is the view factor, which is assumed to be 0.02 according the geometry of the setup. In ideal situation (e.g. the reflection of GNM within thermal radiation range is unity and emission of the emitter within thermal radiation range is zero), $P_{obj}(T_{obj})$ equals to zero. In this condition, the theoretical limit for double-sided radiative cooling at 22 °C is 313.8 W/m2.

**Data availability.** The data that support the findings of this study are available from the corresponding author, Q. Gan, upon request.

**Acknowledgements**

This work was supported by the National Science Foundation (grand no. CBET-1932968 and 1932843).

**Author contributions**

Q.G. conceived the idea and supervised the project. L.Z., H.S., N.Z., J.R. and M.S. executed the experiments. All authors contributed to the analysis of the experimental results and modeling. L.Z., H.S., J.R., M.S., H.Z., Z.Y., B.O. and Q.G. wrote the manuscript. All authors reviewed the manuscript. L.Z and H.S are co-first authors and contributed equally.

**Competing interests**

L.Z., H.S., Z.Y., Q.G. are named as inventors on a provisional patent application pertaining to this work (US 62/844,120). Q.G. and Z.Y. have founded a company, Sunny Clean Water LLC, seeking to commercialize the results reported in this paper.

**Supplementary materials:**

Table S1

Figs. S1 to S8

# Graded nanocomposite metamaterials for a double-sided radiative cooling architecture with a record breaking cooling power density

Lyu Zhou[*,1], Haomin Song[*,1], Nan Zhang[1], Jacob Rada[1], Matthew Singer[1], Huafan Zhang[2], Boon S. Ooi[2], Zongfu Yu[3], Qiaoqiang Gan[1]

1 *Department of Electrical Engineering, The State University of New York at Buffalo, Buffalo, NY 14260, USA.*

2 *Department of Electrical and Computer Engineering, University of Wisconsin, Madison, Wisconsin 53705, USA*

3 *KAUST Nanophotonics Lab, King Abdullah University of Science and Technology, Thuwal 23955-6900, Saudi Arabia*

* These authors contribute equally to this work.

† Email: qqgan@buffalo.edu

**Supplementary Table 1. Atomic percentage of all elements obtained from EDS analysis**

| Layer | C | O | Al | Si | S | Ti | Ag | others | Ag/Si ratio |
|---|---|---|---|---|---|---|---|---|---|
| 2 | 11.54 | 45.12 | 9.37 | 6.03 | 1.06 | 3.83 | 23.04 | 0 | 3.82089552 |
| 3 | 10.26 | 45.69 | 8.33 | 6.79 | 1.1 | 4.02 | 23.81 | 0 | 3.50662739 |
| 4 | 11.05 | 43.54 | 7.96 | 7.26 | 0.97 | 3.87 | 25.36 | 0 | 3.49311295 |
| 5 | 9.83 | 49.95 | 7.45 | 6.89 | 0.85 | 3.4 | 21.08 | 0.54 | 3.05950653 |
| 6 | 9.26 | 51.47 | 7.83 | 6.56 | 0.98 | 3.43 | 19.88 | 0.58 | 3.0304878 |
| 7 | 8.66 | 53.21 | 7.27 | 6.9 | 0.91 | 3.47 | 19.04 | 0.54 | 2.75942029 |
| 8 | 11.01 | 51.8 | 8.12 | 6.63 | 0.94 | 3.93 | 17.57 | 0 | 2.65007541 |
| 9 | 8.64 | 56.11 | 6.9 | 6.84 | 1.04 | 2.99 | 17.47 | 0 | 2.55409357 |

**Supplementary Figure 1. Infrared images of a GNM film and a black aluminum foil**

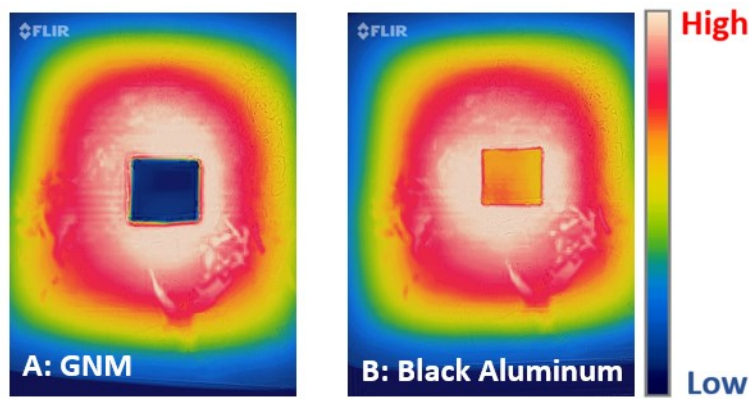

**Fig. S1.** Infrared images of (A) a GNM film and (B) black aluminum foil on a heating stage. Two samples were placed on a hot plate at a temperature of 250 ºC. The infrared images were taken by a thermal camera (FLIR One Pro). To ensure accurate measurement, the temperature of the hot plate, the GNM film and the black aluminum foil was measured with a thermal probe. Although the actual temperature of these two samples are similar, the relative darkness of the GNM sample shown in the thermal image can be credited to the low thermal emissivity.

**Supplementary Figure 2. Optimization of the angle for the GNM plate in the double-sided system**

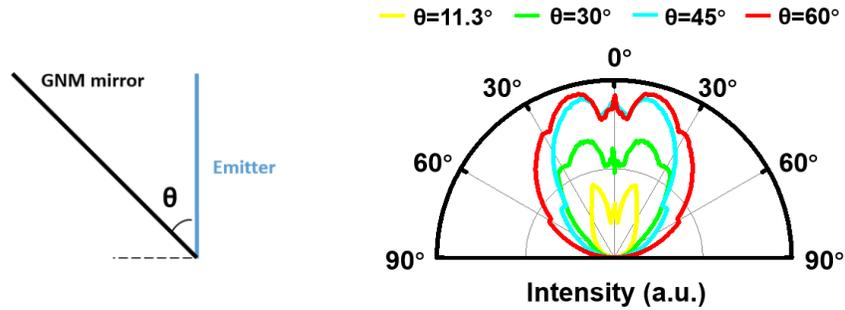

**Fig. S2.** The angle, θ, between the emitter and the GNM mirror (left panel) can be tuned to optimize the thermal reflection, i.e., thermal emission of the system. The modeled angle-dependent emission intensity distribution is plotted in the right panel. One can see that the emission intensities for θ = 60º (red curve) and θ = 45º (cyan curve) are substantially higher than that for θ = 30º (green curve) and θ = 11.3º (yellow curve). Although an angle of 60° yields the highest intensity, we opted for θ = 45º for the following reasons: with a larger angle (θ = 60º), (1) one will need GNM plates with a larger area to efficiently reflect the thermal emission from the vertically aligned emitter; (2) the emitter is exposed to direct sunlight for longer time, due to the larger opening between the emitter and the GNM mirror.

**Supplementary Figure 3. The absorption spectrum of the black aluminum foil**

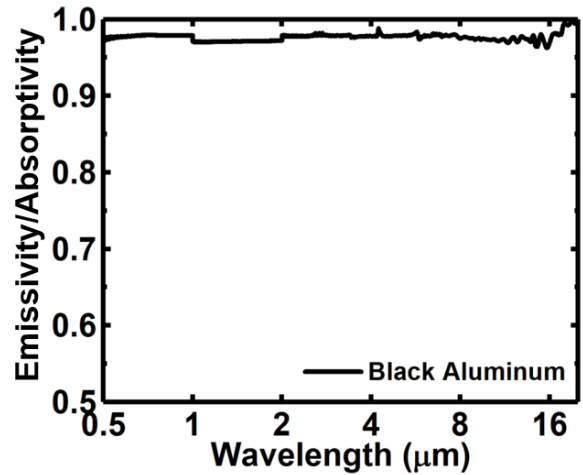

**Fig. S3.** Absorption spectrum of the black aluminum foil.

**Supplementary Figure 4. Atmospheric transmittances under varying water column condition.**

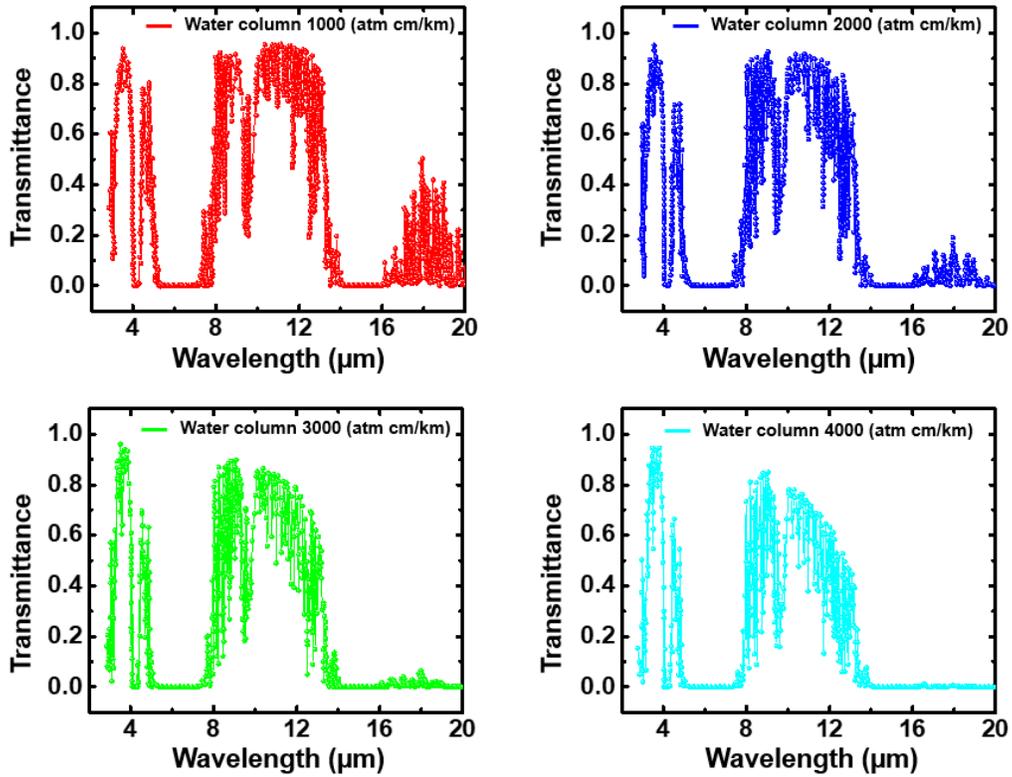

**Fig. S4.** Atmospheric transmission in different water columns. The atmospheric transmittances are obtained from MODTRAN for mid-latitude regions during summer months [1]. The figures show the atmospheric transmittances under different water column (precipitable water) conditions: (a) 1000 atm cm/km, (b) 2000 atm cm/km, (c) 3000 atm cm/km and (d) 4000 atm cm/km. As the water column increased, the transmittance within the atmospheric window decreased significantly, thus reducing the net cooling power.

**Supplementary Figure 5. Calculation of the solar elevation angle**

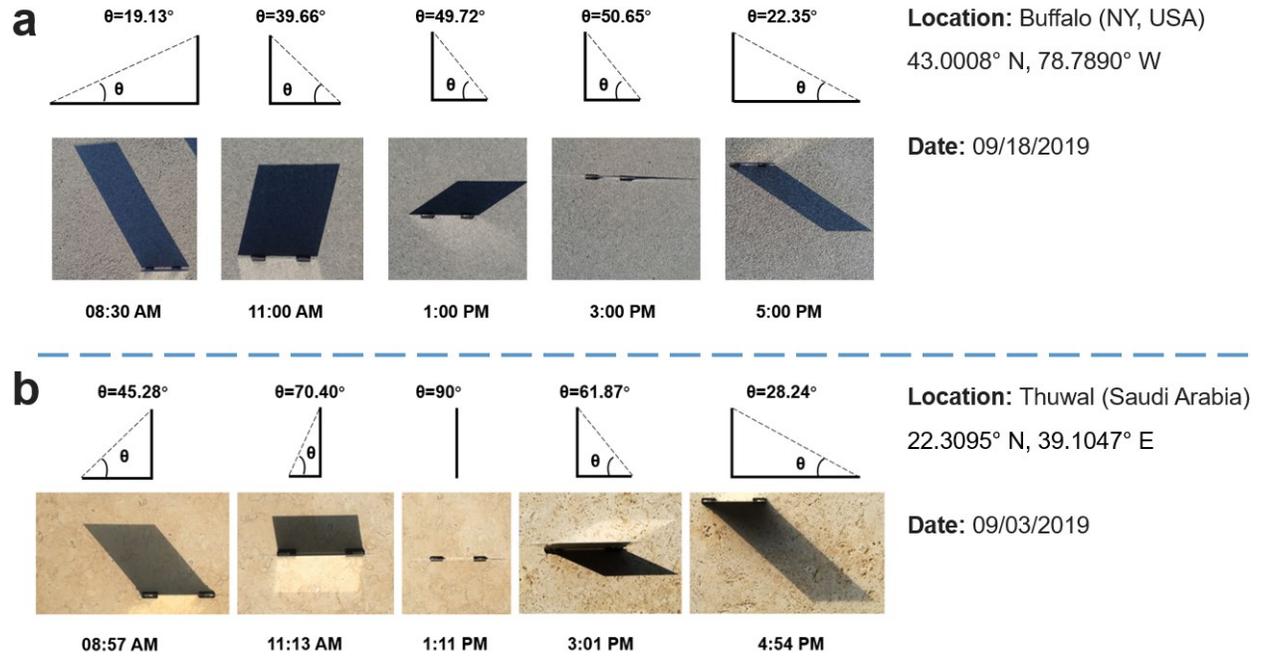

**Fig. S5.** Photos of shadows projected by a 12 cm × 12 cm aluminum plate at different times in (a) Buffalo, USA and (b) Thuwal, Saudi Arabia. The upper panel calculates the elevation angle based on the projected shadow dimension. Either in Buffalo or Thuwal, at certain moment (3:00 pm in Buffalo and 1:11 pm in Thuwal), the planar solar shelter cannot protect the thermal emitter from direct exposure to the sun.

**Supplementary Figure 6. 40-hour continuous radiative cooling test at Buffalo on August 1$^{st}$ to 3$^{rd}$, 2019**

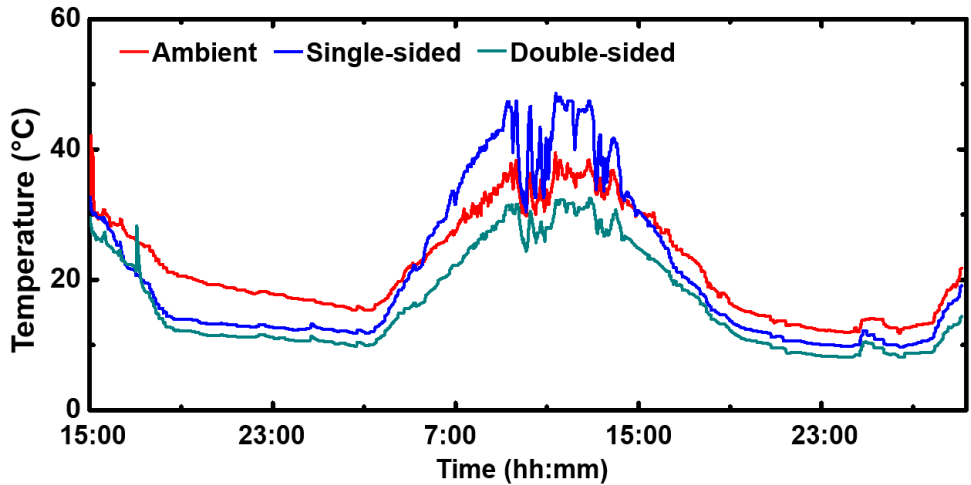

**Fig. S6.** Outdoor radiative cooling test on August 1$^{st}$ to 3$^{rd}$, 2019. The humidity is ~40% in the day and ~98% at night. One can see that the double-sided system continuously outperforms the single-sided system throughout the entire test.

**Supplementary Reference**